# Temperature dependent anisotropic charge carrier mobility limited by ionized impurity scattering in thin-layer black phosphorus

Yue Liu[*] and P. Paul Ruden

*Department of Electrical and Computer Engineering, University of Minnesota, Minneapolis, Minnesota, 55455, USA*

**ABSTRACT**

Anisotropic charge carrier transport in black phosphorus limited by ionized impurity scattering at finite temperature is explored theoretically. The anisotropic electronic structure enters the calculation for the polarizability (screening), the momentum relaxation time, and the mobility. For finite temperature, scattering is not limited to the Fermi surface and the polarizability is temperature dependent. The impact of screening is investigated in detail with its dependence on carrier density and temperature. Competing with the thermal excitation effects, the temperature dependence of the polarizability is found to dominate for $T < 100K$. As a result, the charge carrier mobility slowly decreases with increasing temperature. The weak temperature dependence of the mobility and its anisotropy ratio of $1.9 - 3.2$ agree with published experimental data.

*Author to whom correspondence should be addressed: liux1387@umn.edu

I. INTRODUCTION

Much as graphene and the transition metal dichalcogenites (TMDs), black phosphorus (BP), an allotrope of phosphorus, is of considerable interest due to its two-dimensional (2D) nature [1-4]. With measured charge carrier mobilities of up to $1000 cm^2/Vs$ [5, 6], a direct tunable band gap of $0.3 - 2.0 eV$ depending on the number of monolayers [7-10], and field effect transistor $I_{on}/I_{off}$ ratios on the order of $10^5$ [2, 11-13], BP is a promising candidate for future electronic devices and for certain optoelectronic applications. Moreover, the puckered crystal structure of BP leads to properties with strong in-plane anisotropy, including carrier mobility and thermal conductivity [2, 5, 14-17], as well as linear dichroism in optical absorption [8-10, 18]. These anisotropies make BP special among the 2D materials of interest.

There have been several works, including one by the authors, focusing on the BP charge carrier mobility and its anisotropy [19-21]. While electron-phonon interaction, which is weak at low temperature, is the dominant scattering mechanism in graphene [24], BP samples are believed to have significant concentrations of defects, suggesting that ionized impurity scattering may be an important scattering mechanism, particularly at low temperature. In this work, we focus on the temperature dependent anisotropic linear transport properties of BP limited by scattering due to charge centers either in the BP layer or off the BP layer in the substrate. The Coulomb interaction between the charge centers and the charge carriers in the BP layer is screened, and that screening is treated in temperature dependent mean field approximation. Throughout, the anisotropy of the electronic structure is taken into account, avoiding approximations that are explicitly or implicitly based on isotropy [9, 19, 20]. We consider ionized impurity scattering as the limiting scattering mechanism and use it as a demonstration vehicle, however the basic approach can be readily extended to other mechanisms such as neutral impurity scattering and (quasi-elastic) phonon scattering. The theory presented in this work could be applied to analyze few-layer BP, where the out-of-plane confinement is quantum limited and single sub-band condition is still valid. Additionally, we are motived by recent experimental measurements indicating that the BP mobility has a small temperature dependence, weaker than $T^{-\frac{1}{2}}$ for $T < 100K$, which may not be explained by phonon scattering. Ionized impurity scattering is furthermore an important mechanism for devices in which BP may be employed as the active material, such as ultra-thin field effect transistors, for which TMDs are also being explored [25-27].

In our previous work, the BP hole mobility anisotropy ratio assuming a spatially uniform distribution of charge centers was found to be 3-4 for zero temperature and a wide range of carrier densities, which is somewhat larger than the ratio of 1.5-2 obtained from experimental measurements at $50 - 100K$. We expect the calculated anisotropy ratio for $T > 0$ to be smaller than for $T = 0$. The non-zero temperature not only affects the charge carrier mobility, it also impacts the screening of the charge carriers.

This paper is organized as follows. After revisiting the Boltzmann transport equation result for the anisotropic momentum relaxation time in linear response and the BP energy dispersion, we explore the impact of non-zero temperature on screening, the relaxation time, and the charge carrier mobility in Section II. Section III presents a brief summary and conclusions.

## II. MODEL DEVELOPMENT AND RESULTS

### A. Anisotropic momentum relaxation time

In linear response, the inverse momentum relaxation time for anisotropic materials can be derived from Boltzmann's transport equation and written as, [21,28]

$$\frac{1}{\tau_m(\hat{\xi}, \vec{k}_i)} = \frac{1}{(2\pi)^2} \int_{all\ \vec{k}_j} d\vec{k}_j\, P_{\vec{k}_i,\vec{k}_j} \left\{ 1 - \frac{[\hat{\xi} \cdot \vec{v}(\vec{k}_j)] \tau_m(\hat{\xi}, \vec{k}_j)}{[\hat{\xi} \cdot \vec{v}(\vec{k}_i)] \tau_m(\hat{\xi}, \vec{k}_i)} \right\} \tag{1}$$

where $\hat{\xi}$ is a unit vector in the direction of the applied electric field. The transition rate, $P_{\vec{k}_i,\vec{k}_j}$, for elastic scattering between the incoming state $|\vec{k}_i\rangle$ and the outgoing state $|\vec{k}_j\rangle$ can be expressed by Fermi's golden rule as,

$$P_{\vec{k}_i,\vec{k}_j} = \frac{2\pi}{\hbar} |\langle \vec{k}_j | H | \vec{k}_i \rangle|^2 n_{imp} \delta[E(\vec{k}_i) - E(\vec{k}_j)] \tag{2}$$

where $H$ is the interaction Hamiltonian. For ionized impurity scattering, $H$ represents the screened Coulomb interaction and it will be discussed below. Due to the anisotropic effective mass $m_{xx} \neq m_{yy}$ for $\hat{x}$ (armchair) and $\hat{y}$ (zigzag) directions, the in-plane constant energy surface for both electrons and holes in BP is ellipsoidal,

$$E(\vec{k}) = \frac{\hbar^2}{2} \left( \frac{1}{m_{xx}} k_x^2 + \frac{1}{m_{yy}} k_y^2 \right) \tag{3}$$

with either the conduction band minimum or the valence band maximum at Γ defined as the origin of the energy scale. $\vec{v}(\vec{k}) = \frac{1}{\hbar}\nabla_{\vec{k}}E(\vec{k})$ is the group velocity. If the out-of-plane confinement is quantum limited and the single sub-band model is valid, the ellipsoidal dispersion, $E(\vec{k})$, may be applied for few-layer BP. We consider a model (suitable to represent the common back-gated structure) in which a thin BP layer is placed on a SiO$_2$ insulating substrate [21]. The carrier density can be tuned externally by an applied gate voltage.

For zero temperature, elastic scattering is limited to the Fermi surface and thus $E(\vec{k}) = E_F$. The implicit integral equation for $\tau_m(\hat{\xi}, \vec{k}_i)$, Eq. (1), can be solved following the method of reference [21]. One finds that the momentum relaxation time depends both on the direction of the electric field, $\hat{\xi}$, and on the angle of the incoming state $\vec{k}_i$ (or $\theta_i$). (For isotropic materials, the momentum relaxation time depends only on $|\vec{k}_i|$).

At non-zero temperatures, the scattering is still an elastic process but no longer limited to $E_F$. Instead, it occurs in the vicinity of $E_F$ following a distribution or sampling function $\left(-\frac{\partial f}{\partial E}\right) = \frac{1}{k_BT}f(E) \cdot [1 - f(E)]$, where $f(E)$ is the Fermi-Dirac distribution. Of course, $\left(-\frac{\partial f}{\partial E}\right)$ reduces to a delta function at $E_F$ in the zero temperature limit. As $T$ increases, the distribution $f(E) \cdot [1 - f(E)]$ spreads. Thus, for $T > 0$, we need to consider a range of energies rather than a single (Fermi) energy. In the following discussion, we write the $\vec{k}_i$ dependence of the relaxation time as a function of energy $E$ and incoming wave vector angle $\theta_i$, i.e. $\tau_m(\hat{\xi}, \vec{k}_i) \equiv \tau_m(\hat{\xi}, E, \theta_i)$.

The Fermi energy itself is also a function of temperature, i.e. $E_F(T) = k_BT \ln\left[\exp\left(\frac{n}{g_{2D}k_BT}\right) - 1\right]$, where $n$ is the 2D carrier density and $g_{2D} = \frac{m_d}{\pi\hbar^2}$ is the 2D density of states (DOS) of the BP layer, with DOS effective mass $m_d = \sqrt{m_{xx}m_{yy}}$. In the following numerical computations, we consider the energy regions $[E_F(T) - 5k_BT < E < E_F(T) + 5k_BT]$ for $E_F(T) - 5k_BT > 0$ and $[0 < E < 10k_BT]$ for $E_F(T) - 5k_BT \leq 0$, since there are no available states for energy less than zero (i.e. below conduction band minimum or above the valence band maximum). The total $10k_BT$ energy regions cover at least 97% of $f(E) \cdot [1 - f(E)]$.

The methods and the discussion in this section are general and can be applied to any elastic scattering mechanism. Ionized impurity scattering is considered later in this work as an example. In addition to the energy, $E$, there is another term in $\tau_m(\hat{\xi}, E, \theta_i)$ that introduces temperature dependence: the polarizibility/ screening examined next.

The matrix elements $|\langle \vec{k}_j|H|\vec{k}_i\rangle|$ for ionized impurity scattering can be written as, [29, 30]

$$|\langle \vec{k}_j|H|\vec{k}_i\rangle| = \frac{2\pi e^2 e^{-qd}}{q\kappa + 2\pi e^2 \Pi(\vec{q},T)} = \frac{2\pi e^2}{\kappa} \frac{e^{-qd}}{q + q_s} \qquad (4)$$

where $\vec{q} = \vec{k}_j - \vec{k}_i$ is the wave vector transferred during scattering, $d$ is the impurity distance from the BP plane, and $\kappa$ is the effective dielectric constant. ($d = 1nm$ is used in the numerical calculations unless stated otherwise; $\kappa = 2.5$ is used for air/SiO$_2$ half spaces). $\Pi(\vec{q},T)$ is the temperature dependent, anisotropic polarizability of the mobile charge carriers in BP. Analogous to the transferred wave vector, $q = |\vec{q}| = \sqrt{q_x^2 + q_y^2}$, indicating scattering strength, the effect of the polarizability is represented by $q_s(\vec{q},T) = \frac{2\pi e^2}{\kappa}\Pi(\vec{q},T)$. For zero temperature, $q \leq 2k_{Fy}$, where $k_{Fy}$ is the $\hat{y}$ direction (heavy effective mass) Fermi wave vector; while for finite temperature, $q$ can be larger than $2k_{Fy}$ because scattering is not limited to the Fermi surface. Linear mean field screening theory is used [31]. For zero temperature, the polarizability has the closed form $\Pi(\vec{q},0) = g_{2D}Re\left[1 - \sqrt{1 - \frac{8\mu/\hbar^2}{q_x^2/m_{xx} + q_y^2/m_{yy}}}\right]$, where $\mu = E_F(0) = \frac{\hbar^2 \pi n}{m_d}$ is the zero temperature Fermi energy [32]. Using Maldague's formula [30, 33], the temperature dependent polarizability $\Pi(\vec{q},T)$ can be written as,

$$\Pi(\vec{q},T) = \int_0^\infty d\mu \frac{\Pi(\vec{q},0)}{4k_B T cosh^2\left(\frac{E_F(T) - \mu}{2k_B T}\right)} \qquad (5)$$

To assist in comparing the magnitudes of $q$ and $q_s$ in the dominator of Eq. (4) with their dependence on carrier density and temperature, the $q_x$ and $q_y$ axes, as well as the screening term $q_s$ are normalized by $k_{Fy}$, which is determined by $E_F$ for a given temperature. The normalized $q_s$ with their dependence on $q_x$ and $q_y$ are shown using the plots of Fig. 1. The magnitude of $q_s/k_{Fy}$ is represented by colors. The dashed white curves correspond to the energy regions of interest ($[E_F(T) - 5k_B T, E_F(T) + 5k_B T]$ or $[0, 10k_B T]$). It is apparent that the screening effects become weaker as temperature increases, but they also depend on the carrier density [34]. For $T = 0K$, $q_s$ is a constant ($\Pi(\vec{q},0) = g_{2D}$ for $q_x \leq 2k_{Fx}$, $q_y \leq 2k_{Fy}$), and the corresponding energy regions reduce to a single curve, $E_F(0)$. For finite temperature, $q_s$ decreases as $q$ increases in the entire $\vec{q}$-plane. The corresponding energy regions display both upper and lower bounds in Fig. 1 (e) and (f), or only upper bounds for Fig. 1 (b) and (c), since these lower bounds are at zero energy.

Comparing constant carrier density cases (Fig. 1(a)-(c) with $n = 2 \times 10^{12} cm^{-2}$ or (d)-(f) with $n = 10^{13} cm^{-2}$), the screening becomes less effective as $T$ increases, for to two reasons. First, $q_s$ itself becomes smaller as $T$ increases [34]. Second, the energy regions, limited by the sampling function $f(E) \cdot [1 - f(E)]$, become wider for higher temperature. When the temperature is low, the energy regions are narrow and mostly include the range of strong screening; while for high temperature, the energy regions are wide and include more of the weak screening range.

Comparing the same temperature cases (Fig. 1 (a) and (d) with $T = 0K$, (b) and (e) with $T = 40K$ or (c) and (f) with $T = 100K$), it is worthwhile to note that although the normalized $q_s/k_{Fy}$ assumes larger values for smaller $n$, one finds that $q_s$ itself increases as $n$ increases for $T > 0K$. This is the expected increase in screening with increasing carrier density. (Except for $T = 0K$, where $q_s = \frac{2\pi e^2}{\kappa} g_{2D}$, independent of $n$) Furthermore, $q_s$ decreases faster as $T$ increases for smaller carrier density, therefore the scattering is more affected by temperature for smaller $n$.

We proceed to calculate the momentum relaxation time for applied electric fields along the $\hat{x}$ (armchair) and $\hat{y}$ (zigzag) directions. The hole effective masses in thin BP can vary from $m_{xx} = 0.15 m_0$ and $m_{yy}$ larger than $3.5 m_0$ in monolayers, to $m_{xx} = 0.04 m_0$ and $m_{yy} = 0.9 m_0$ [35,36] in bulk, depending on the calculation or characterization methods. In this work, we begin with the set of effective masses $m_{xx} = 0.15 m_0$ and $m_{yy} = 1.0 m_0$ (same as used in previous work [21]) and discuss the influence of effective masses towards the end of the paper. Hole effective masses are used in the numerical calculation since most BP devices fabricated recently are p-channel field effect transistors [2,11,13,37]. Following the approaches discussed in previous work [21], $\tau_m$ can be calculated numerically from Eq. (1), and it is plotted in Fig. 2. $N_\theta = 400$ points are used for the angle mesh and $N_E = 100$ points for the energy mesh. An impurity density of $n_{imp} = 10^{12} cm^{-2}$ is assumed. Fig. 2 (a) and (b), are for different directions of the electric field, $\hat{x}$ and $\hat{y}$. For Fig. 2 (a) and (c), the carrier densities are different. In Fig. 2 (a) and (d), the carrier density and field direction are the same, but the temperatures are different. The momentum relaxation time surfaces can have various shapes subject to the different conditions (direction of electric field, carrier density, and temperature). This is due to the anisotropic electronic structure of BP, which is taken into account through the implicit integral equation (Eq. (1)). Without anisotropy, $\tau_m$ would only depend on $E$ (or $|\vec{k}_i|$).

It is helpful to obtain a better understanding of $\tau_m(\hat{\xi}, E, \theta_i)$ since it has significant impact on the mobility. The dependence of $\tau_m$ on the angle $\theta$ and on the carrier density $n$ was investigated in previous work [21]. Hence, we focus on the dependence on energy and on temperature. We divide the energy into three ranges in the following discussion, depending on the different effects observed, and frequently refer back to Fig.1 for the discussion of the energy dependence. To analyze the latter and eliminate the angular dependence, a $\bar{\tau}_m(\hat{\xi}, E)$, which is the momentum relaxation time averaged over the incident angle, is calculated:

$$\bar{\tau}_m(\hat{\xi}, E) = \frac{1}{2\pi} \int_0^{2\pi} d\theta_i \tau_m(\hat{\xi}, E, \theta_i) \qquad (6)$$

Results are plotted in Fig. 3 for different $T$ and $E$. For convenience, the trend of increasing $T$ is indicated in Fig. 3 (a) and (b), while three energy ranges are labeled in Fig. 3 (c) and (d). For very low temperatures, $\bar{\tau}_m$ has a rather weak dependence on energy regardless of the direction of the electric field and the carrier density. This is due to the energy region of the sampling functions $f(E) \cdot [1 - f(E)]$ being quite narrow for low temperature, and neither $q$ nor $q_s$ change much over that range. For somewhat higher temperatures, one observes that $\bar{\tau}_m$ first slightly increases, then slightly decreases with $E$ (10-20$K$ curves in Fig. 3). This is still due to the narrow $E$ range of the sampling function for the relatively low temperature. The effective screening term $q_s$ in Fig. 1 is essentially constant for the energy range. But as the energy increases, $q$ in the denominator of Eq. (4) increases. Consequently, the scattering matrix element $|\langle \vec{k}_j | H | \vec{k}_i \rangle|$ decreases and $\bar{\tau}_m$ increases. This is the "first range" of $E$. As $E$ continues to increase and becomes larger than $E_F$, $q_s$ decreases rapidly. If $q_s > q$, the effect of decreasing $q_s$ dominates. Thus $|\langle \vec{k}_j | H | \vec{k}_i \rangle|$ increases and $\bar{\tau}_m$ decreases. This is the "second range" of $E$. Similar behavior has been found in the scattering rate (i.e. the reciprocal of $\tau_m$) for isotropic silicon inversion layers [34].

For relatively high temperatures (e.g. $T$ =70-100$K$), the behavior of $\bar{\tau}_m$ is more complicated since it varies with $\hat{x}$ and $\hat{y}$ directions and the carrier densities. For high temperature, the energy range of the sampling function $f(E) \cdot [1 - f(E)]$ is large. Furthermore, as $E$ becomes large, the screening term $q_s$ becomes so small that $q$ again dominates. Thus $\bar{\tau}_m$ once again increases as $E$ increases, such as shown by the 40$K$ curve in Fig. 3 (c) or the 100$K$ curve in Fig. 3 (d). This is the "third range" of $E$. Under certain circumstances, the $q_s$-dominated second range is so small that it is masked by the "first" and "third ranges", hence $\bar{\tau}_m$ seems to increase monotonically as $E$ increases. For instance, the high temperature curves (70-100$K$) in Fig. 3 (a) and (c), increase with $E$ monotonically, but they increase slowly within the moderate energy range. The difference between $\hat{x}$ and $\hat{y}$ directions is due to the magnitudes of $q_x$ and $q_y$. Since $q_y$ is relatively large, it is more likely that a "third range" appears for the $\hat{y}$ direction, while for the $\hat{x}$ direction, the "third range" may only exist at even higher $T$. Since the energy regions covered by $f(E) \cdot$

$[1 - f(E)]$ for $n = 2 \times 10^{12} cm^{-2}$ ($[0, 10k_BT]$) and $n = 10^{13} cm^{-2}$ ($[E_F(T) - 5k_BT, E_F(T) + 5k_BT]$) are not the same, the lower and upper bounds on the energy axes in Figs. 3 (a) and (b), (c) and (d) are different. For other situations (e.g. different dielectric constant or device structure), the detailed shapes of the curves may change, but the basic behavior is expected to be similar.

### B. Mobility

With the momentum relaxation time $\tau_m(\hat{\xi}, E, \theta_i)$ calculated, we can proceed to calculate the temperature dependent anisotropic mobility as follows (taking the $\hat{x}$ direction as an example), [28]

$$\mu_{xx} = \frac{g_s e}{(2\pi)^2 n \hbar^2} \int_0^{2\pi} d\theta_i \int_0^{\infty} dE \frac{k_i(E, \theta_i)}{\left|\frac{\partial E}{\partial k_i}\right|_{\vec{k}_i(E,\theta_i)}} \left(\frac{\partial E}{\partial k_{ix}}\right)^2 \tau_m(\hat{x}, E, \theta_i) \left(-\frac{\partial f}{\partial E}\right) \quad (7)$$

where $g_s = 2$ is the spin degeneracy, $k_i = |\vec{k}_i|$, $k_{ix} = k_i \sin\theta_i$. The result for $\mu_{yy}$ is analogous. $\mu_{xx}$ and $\mu_{yy}$ are plotted in Fig. 4 as functions of $T$.

There are two competing mechanisms determining the temperature dependence of the mobility. At low temperature, screening dominates: as $T$ increases, the screening term, $q_s$, becomes smaller (Fig. 1) and the scattering matrix element $|\langle \vec{k}_j | H | \vec{k}_i \rangle|$ in Eq. 4 becomes larger; more scattering events occur and thus the relaxation time and mobility decrease. At high temperature, thermal excitation dominates: the carriers have greater thermal energy, and the mobility increases as $T$ increases [38]. For BP, the transition between the two regimes, screening dominating and thermal excitation dominating, depends on several parameters, such as the carrier density and the direction of the electric field, but it generally is around $100$-$120 K$. Therefore, in the temperature range of interest for ionized impurity scattering, the mobility tends to decrease with increasing temperature as shown in Fig. 4, due to the dominance of the temperature dependent screening. (Without considering temperature dependence for screening by setting $q_s = \frac{2\pi e^2}{\kappa} g_{2D}$ as a constant (zero temperature screening case), the calculated $\mu_{xx}$ and $\mu_{yy}$ can be found to increase as temperature increases for all $T$.)

We also find that the rate of decrease with increasing temperature is smaller for $\mu_{yy}$ than for $\mu_{xx}$. This is consistent with the results shown in Fig 3 for $\bar{\tau}_m(\hat{y}, E)$, which shows a more pronounced "third range" than $\bar{\tau}_m(\hat{x}, E)$. The rate of mobility decrease slows as $T$ continues to increase because the effect of the increasing thermal energy of the charge carriers becomes comparable to the screening effect. By curve fitting, a relation $\mu \propto T^{-\alpha}$ is found. Taking $n = 2\times10^{12} cm^{-2}$ as an example, one obtains $\alpha = 0.27$ for $\mu_{xx}$ and $\alpha = 0.12$ for $\mu_{yy}$, the weak dependence is similar to the experimental data for $T < 100K$ [14,15]. For higher temperatures ($T > 140K$), $\mu_{yy}$ may increase as $T$ increases for certain carrier densities, but that temperature region is likely to be beyond the range of mobilities that are limited by ionized impurity scattering.

As previously discussed, the mobility increases as the carrier density, $n$, increases. Mobility has a stronger dependence on $T$ for smaller $n$ because screening is more sensitive to temperature for lower carrier density, as shown in Fig. 1. Overall the mobility has a weak dependence on temperature for $T < 100K$. The calculated $\mu_{xx}$ and $\mu_{yy}$ (Fig. 4) are considerably larger than currently reported experimental BP hole mobilities, indicating that the impurity concentration of the samples fabricated may well be larger than the assumed value of $n_{imp} = 10^{12} cm^{-2}$. To eliminate the effect of $n_{imp}$, the anisotropy ratio, $\frac{\mu_{xx}}{\mu_{yy}}$, may be seen as a better quantity for comparison, and its temperature dependence is plotted in Fig. 5. Since $\mu_{xx}$ decreases more strongly than $\mu_{yy}$ with increasing temperature, the anisotropy ratio decreases as $T$ increases. An anisotropy ratio of 1.9-3.2 can be found using hole effective masses at $T\sim 50K$ and carrier densities from $2\times 10^{12} cm^{-2}$ to $10^{13} cm^{-2}$, which is smaller than the previous zero temperature result [21] and closer to the experimental results for few-layer BP : 1.6-4 [14, 15].

Moreover, as $n$ increases, $\frac{\mu_{xx}}{\mu_{yy}}$ decreases. Therefore, BP electronic devices operating at low temperature and small carrier density should display stronger anisotropy. The anisotropy ratio calculated using other parameters such as zero impurity distance, uniform impurity distribution model (averaging the effect of $d$ in a range $0 < d < 300nm$, while keeping the total $n_{imp} = 10^{12} cm^{-2}$ fixed [21]) and results for electrons with effective mass parameters $m_{xx} = 0.15m_0$ and $m_{yy} = 0.7m_0$ are plotted in Fig. 5 (b). Trivially, a reduced effective mass ratio leads to a smaller $\frac{\mu_{xx}}{\mu_{yy}}$. The anisotropy ratio decreases as $d$ increases (i.e. impurities located further away from BP plane), which also agrees with our zero temperature calculation [21])

### III. SUMMARY AND CONCLUSIONS

To summarize, ionized impurity scattering limited anisotropic charge carrier mobilities in BP are explored with focus on their temperature dependence. The anisotropic electronic structure affects the mobilities through the anisotropic polarizability and momentum relaxation time. For $T < 100K$, the screening is found to dominate the temperature dependence, though this effect becomes weaker as $T$ increases. Overall, the mobility decreases as the temperature increases, but the dependence is relatively weak. The calculated mobility dependence on temperature has the same trend as experimental reports for $T < 100K$, indicating ionized impurity may be the dominant scattering mechanism in this temperature region. An anisotropy ratio, $\frac{\mu_{xx}}{\mu_{yy}}$, of 1.9-3.2 is found for $T \sim 50K$ and a wide range of carrier densities. Larger ratios are obtained at lower temperature and lower carrier density.

The basic approach used for calculating the ionized impurity scattering limited momentum relaxation time and mobility is readily extended to other elastic scattering mechanisms. The tendency of the Coulomb matrix element to decrease with increasing transfer wave vector weakens the anisotropy of the momentum relaxation time. This is particularly apparent when comparing results for different values of the parameter $d$. It implies that scattering mechanisms that are independent of the transfer wave vector display greater anisotropy than ionized impurity scattering. For example, neutral impurity scattering and acoustic phonon scattering (in the quasi-elastic limit) may fall into this category. However, other effects related to anisotropy, e.g. the anisotropy of the phonon dispersion or of the deformation potential will enter. This will be the subject of future work.

## ACKNOWLEDGEMENT

The authors thank Dr. Tony Low at the University of Minnesota for stimulating discussions and for a careful reading of the manuscript. Access to the facilities of the Minnesota Supercomputing Institute for High Performance Computing is gratefully acknowledged.

**References:**

[1] A. Morita, *Appl. Phys. A* **39**, 227 (1986)

[2] H. Liu, A. T. Neal, Z. Zhu, D. Tomanek, and P. D. Ye, *ACS Nano* **8**, 4033 (2014)

[3] J. Qiao, X. Kong, Z.-X. Hu, F. Yang, and W. Ji, *Nat. Commun.* **5**, 4475 (2014)

[4] X. Ling, H. Wang, S. Huang, F. Xia, and M. S. Dresselhaus, *Proc. Natl. Acad. Sci. U.S.A.* **112**, 4523 (2015)

[5] L. Li, Y. Yu, G. J. Ye, Q. Ge, X. Ou, H. Wu, D. Feng, X. H. Chen, and Y. Zhang, *Nat. Nanotechnol.* **9**, 372 (2014)

[6] H. Liu, Y. Du, Y. Deng, and P. D. Ye, *Chem. Soc. Rev.* **44**, 2732 (2015)

[7] A. Castellanos-Gomez, L. Vicarelli, E. Prada, J. O. Island, K. L. Narasimha-Acharya, S. I. Blanter, D. J. Groenendijk, M. Buscema, G. A. Steele, J. V. Alvarez *et al*., *2D Mater.* **1**, 025001 (2014)

[8] T. Low, A. S. Rodin, A. Carvalho, Y. Jiang, H. Wang, F. Xia, and A. H. Castro Neto, *Phys. Rev. B* **90**, 075434 (2014)

[9] J. Qiao, X. Kong, Z.-X. Hu, F. Yang, and W. Ji, *Nat. Commun.* **5**, 4475 (2014)

[10] V. Tran, R. Soklaski, Y. Liang, and L. Yang, *Phys. Rev. B* **89**, 235319 (2014)

[11] N. Haratipour, M. C. Robbins, and S. J. Koester, *IEEE Electron Device Lett.* **36**, 411 (2015)

[12] X. Cao and J. Guo, *IEEE Trans. Electron Devices,* **62**, 659 (2015)

[13] N. Haratipour, S. Namgung, S.-H. Oh, and S. J. Koester, *ACS Nano* **10**, 3791 (2016)

[14] F. Xia, H. Wang, and Y. Jia, *Nat. Commun.* **5**, 4458 (2014)

[15] A. Mishchenko, Y. Cao, G. L. Yu, C. R. Woods, R. V. Gorbachev, K. S. Novoselov, A. K. Geim, and L. S. Levitov, *Nano Letters*. **15**, 6991 (2015)

[16] Z. Luo, J. Maassen, Y.Deng, Y. Du, R. P. Garrelts, M. S. Lundstrom, P. D. Ye, and X. Xu *Nat. Commun.* **6**, 8572 (2015).

[17] J. Zhu, J.-Y. Chen, H. Park, X. Gu, H. Zhang, S. Karthikeyan, N. Wendel, S. A. Campbell, M. Dawber, X. Du, M. Li, J.-P. Wang, R. Yang, and X. Wang, *Adv. Electron. Mater.* **2**, 1600040 (2016)

[18] H. Yuan, X. Liu, F. Afshinmanesh, W. Li, G. Xu, J. Sun, B. Lian, A. G. Curto, G. Ye, Y. Hikita *et al*., *Nat. Nanotechnol.***10**, 707 (2015)

[19] Z. -Y. Ong, G. Zhang, and Y. W. Zhang *J. Appl. Phys.* **116**, 214505 (2014)

[20] B. Liao, J. Zhou, B. Qiu, M. S. Dresselhaus, and G. Chen, *Phys. Rev. B* **91**, 235419 (2015)

[21] Y. Liu, T. Low, and P. P. Ruden, *Phys. Rev. B* **93**, 165402 (2016)

[22] A. N. Rudenko, S. Brener, and M.I. Katsnelson, *Phys. Rev. Lett.* **116**, 246401 (2016)

[23] M. V. Fischetti, and W. G. Vandehberghe, *IEEE International Electron Devices Meeting*, 4.2, (2016)

[24] E. H. Hwang, and S. Das Sarma, *Phys. Rev. B* **79**, 165404 (2009)

[25] H. Fang, S. Chuang, T. C. Chang, K. Takei, T. Takahashi, and A. Javey, *Nano Letters*. **12**, 3788 (2012)

[26] M.-C. Chen, K.-S. Li, L.-J. Li, A.-Y. Lu, M.-Y. Li, Y.-H. Chang, C.-H. Lin, A. B. Sachid, T. Wang, F.-L. Yang, and C. Hu, *IEEE International Electron Devices Meeting*, 32.2, (2016)

[27] A. Pezeshki, S. H. H. Shokouh, P. J. Jeon, I. Shackery, J. S. Kim, I.-K. Oh, S. C. Jun, H. Kim, and S. Im, *ACS Nano* **10**, 1118 (2016)

[28] M. V. Fischetti, Z. Ren, P. M. Solomon, M. Yang, and K. Rim, *J. Appl. Phys.* **94**, 1079 (2003)

[29] D. A. Dahl, and L. J. Sham, *Phys. Rev. B* **16**, 651 (1977)

[30] T. Ando, A. B. Fowler, and F. Stern, *Rev. Mod. Phys.* **54**, 437(1982)

[31] Y. Liu, A. Goswami, F. Liu, D. L. Smith, and P. P. Ruden, *J. Appl. Phys.* **116**, 234301 (2014)

[32] T. Low, R. Roldan, H. Wang, F. Xia, P. Avouris, L. M. Moreno, and F. Guinea, *Phys. Rev. Lett.* **113**, 106802 (2014)

[33] P. F. Maldague, *Surf. Sci.* **73**, 296 (1978)

[34] F. Stern, *Phys. Rev. Lett.* **44**, 1469 (1980)

[35] S. ichiro Narita, S. ichi Terada, S. Mori, K. Muro, Y.Akahama, and S. Endo, *J. Phys. Soc. Jpn*. **52**, 3544 (1983)

[36] Y. Cai, G. Zhang, and Y.-W. Zhang, *Sci. Rep*. **4**, 6677 (2014)

[37] L. Li, M. Engel, D.B. Farmer, S. Han, and H.-S. P. Wong, *ACS Nano* **10**, 4672 (2016)

[38] R. F. Pierret, *Advanced Semiconductor Fundamentals*, 2nd ed. (Modular Series on Solid State Devices, Volume VI, Pearson Education, (2002)

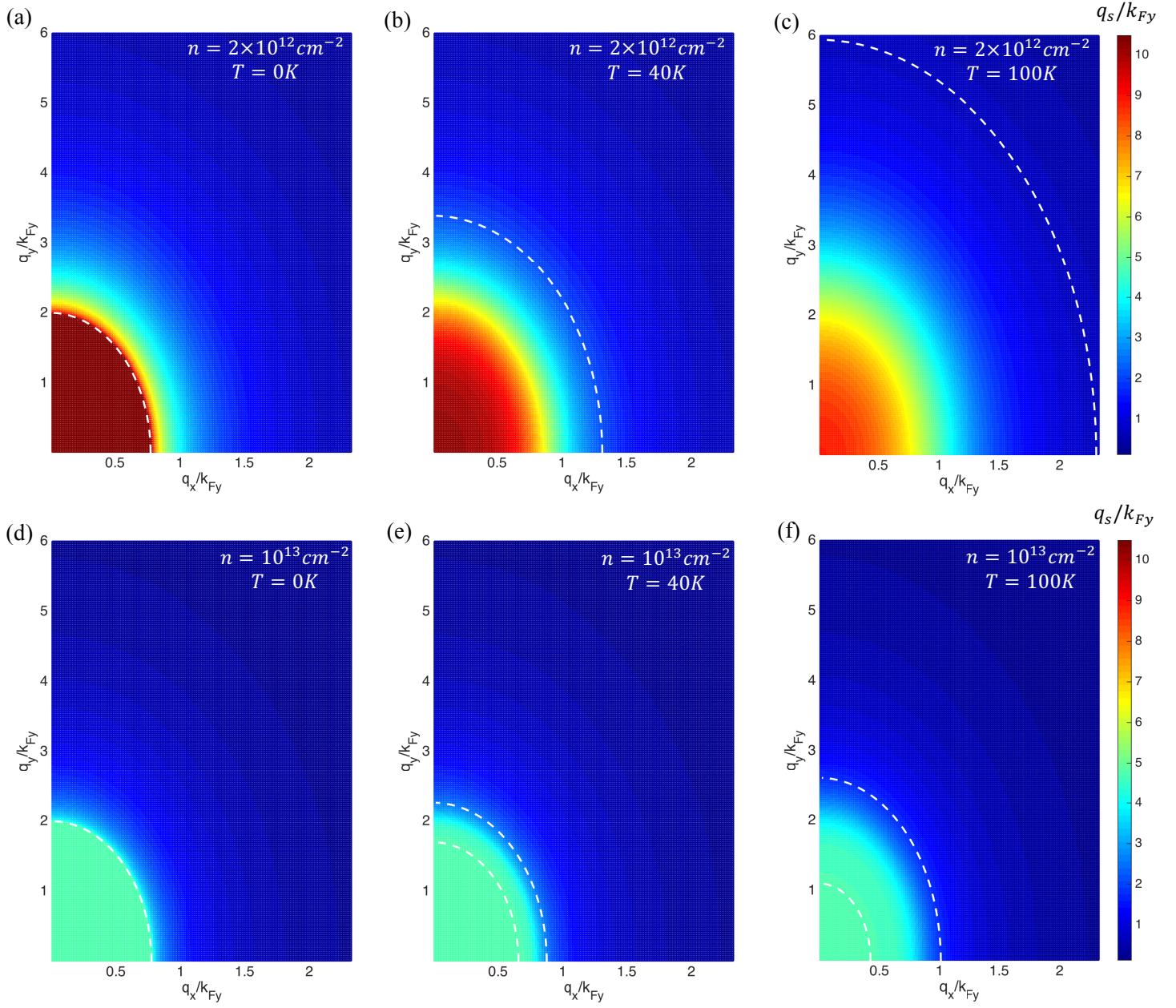

Fig. 1. (Color online) Effective screening term, $q_s$, as a function of $q_x$ and $q_y$ (all normalized to $k_{Fy}$). Dashed white curves correspond to the energy regions as discussed in Section II. A. (a)-(c) $n = 2\times10^{12} cm^{-2}$, with $T = 0, 40, 100 K$. (d)-(f) $n = 10^{13} cm^{-2}$, with $T = 0, 40, 100 K$.

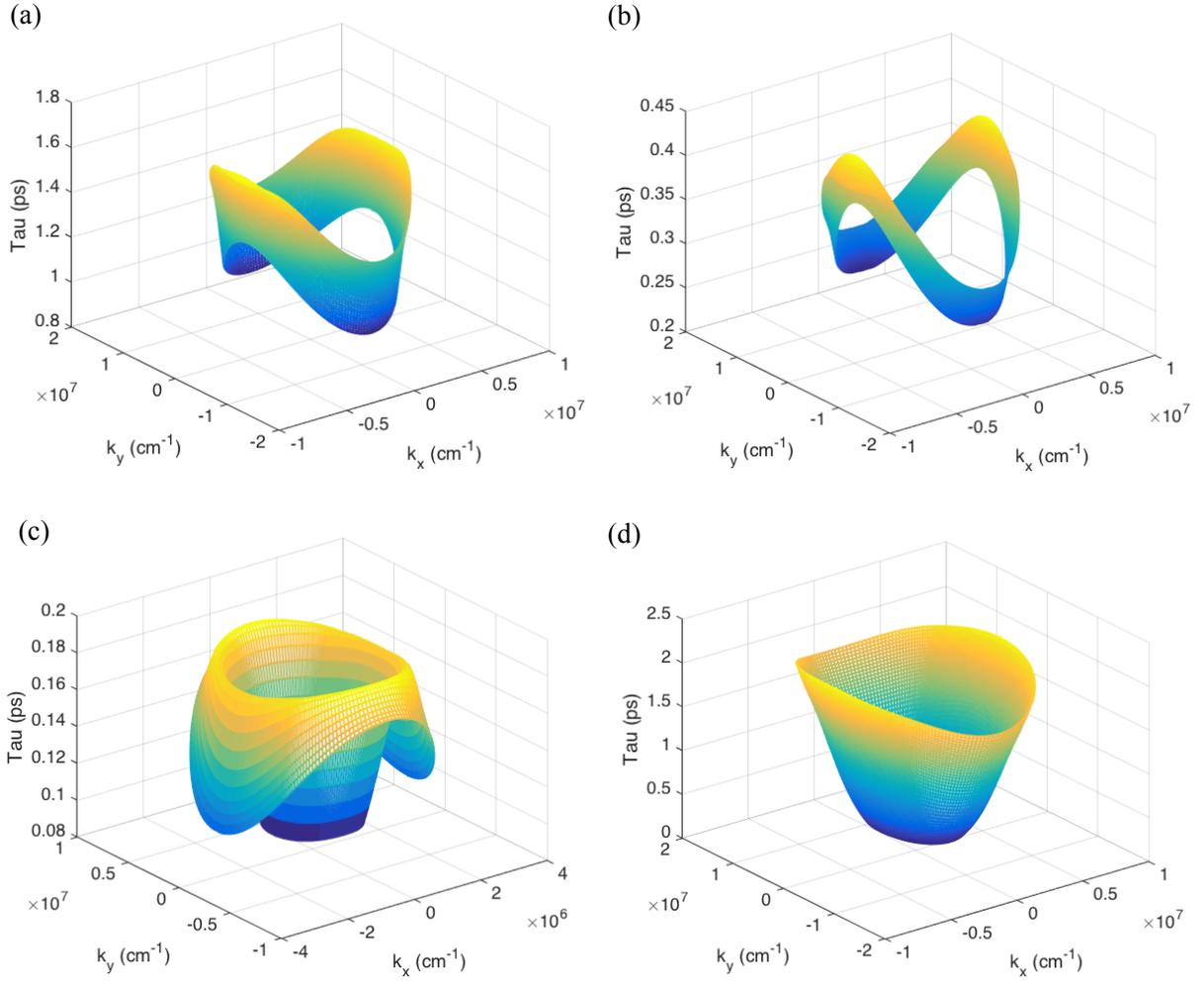

Fig. 2. (Color online) Calculated momentum relaxation time, $\tau_m(\hat{\xi}, E, \theta_i)$, for different directions of the electric field, different carrier densities, and temperatures. (a) $\tau_m(\hat{y}, E, \theta_i)$ with $n = 10^{13} cm^{-2}$, $T = 20K$. (b) $\tau_m(\hat{x}, E, \theta_i)$ with $n = 10^{13} cm^{-2}$, $T = 20K$. (c) $\tau_m(\hat{y}, E, \theta_i)$ with $n = 2\times10^{12} cm^{-2}$, $T = 20K$. (d) $\tau_m(\hat{y}, E, \theta_i)$ with $n = 10^{13} cm^{-2}$, $T = 100K$.

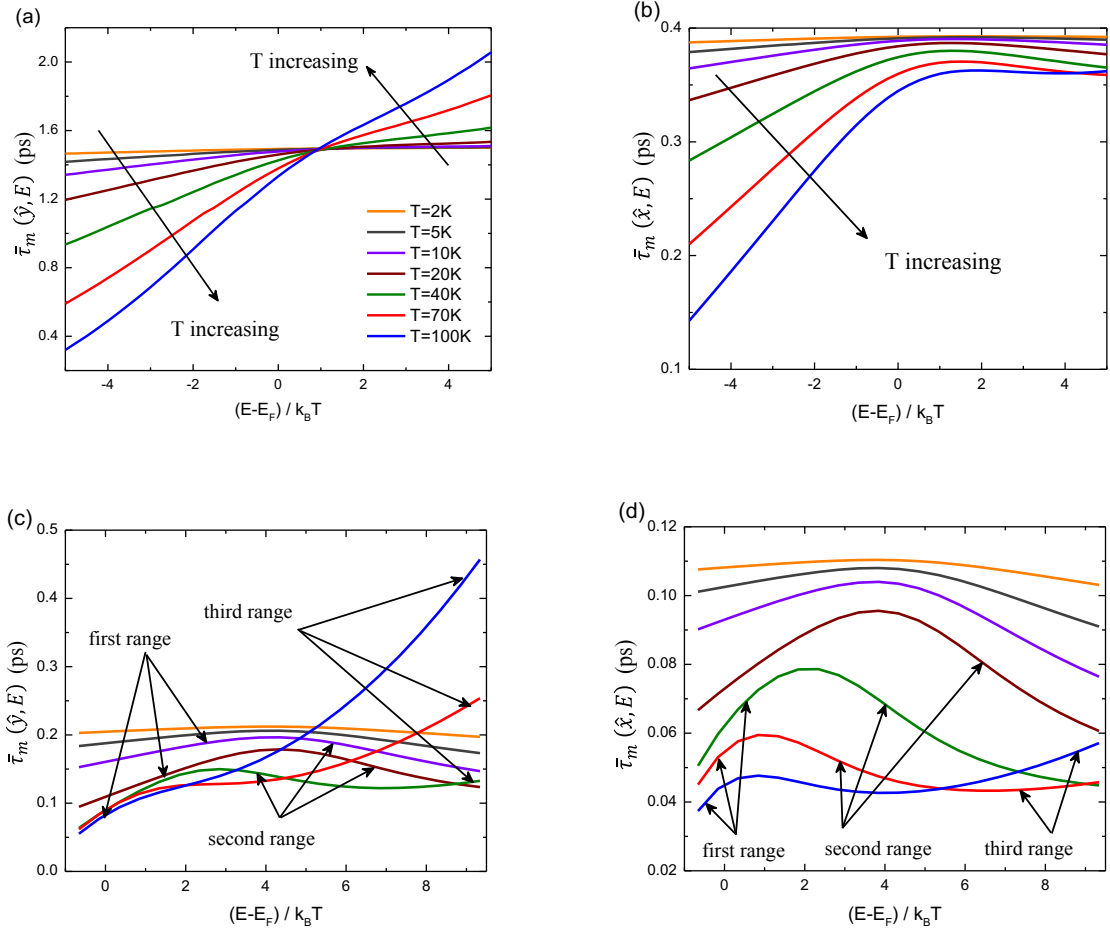

Fig. 3. (Color online) Angle averaged momentum relaxation time, $\bar{\tau}_m(\hat{\xi}, E)$, for different directions of the electric field, differen carrier densities and temperatures. The ranges labelled first, second, and third refer to the different temperature dependences. (a) $\bar{\tau}_m(\hat{y}, E)$ with $n = 10^{13} cm^{-2}$. (b) $\bar{\tau}_m(\hat{x}, E)$ with $n = 10^{13} cm^{-2}$. (c) $\bar{\tau}_m(\hat{y}, E)$ with $n = 2\times10^{12} cm^{-2}$. (d) $\bar{\tau}_m(\hat{x}, E)$ with $n = 2\times10^{12} cm^{-2}$.

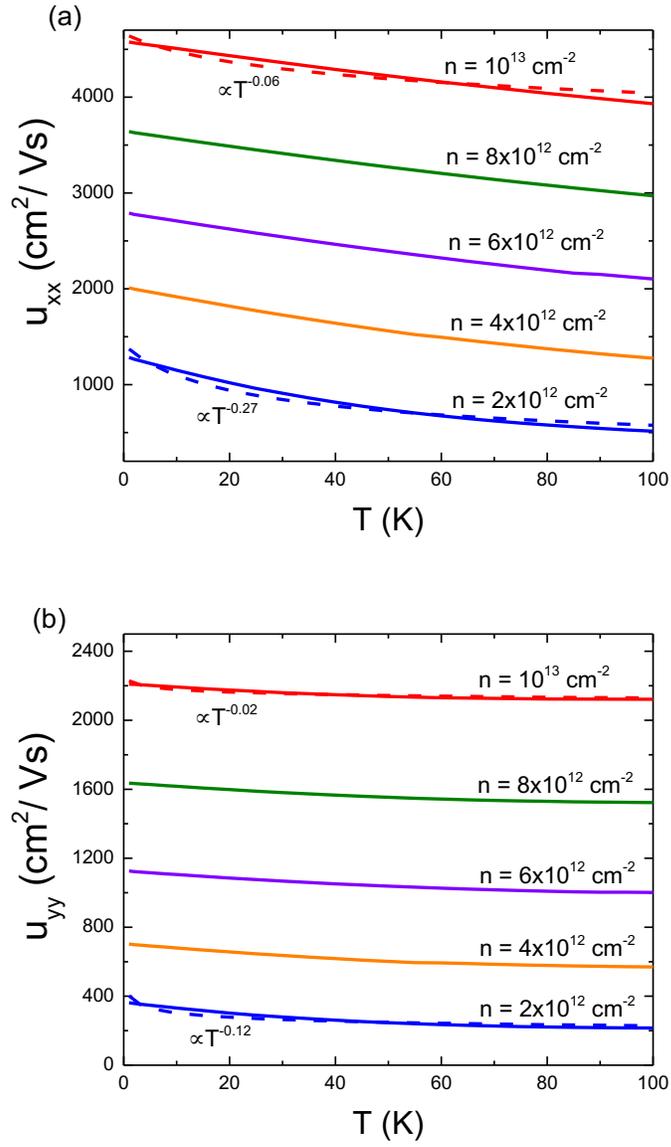

Fig. 4. (Color online) Hole mobility as a function of temperature for several carrier densities. Impurity density $n_{imp} = 10^{12} cm^{-2}$, impurity distance $d = 1nm$, dashed lines are fits of $\mu \propto T^{-\alpha}$. (a) $\mu_{xx}$ (b) $\mu_{yy}$.

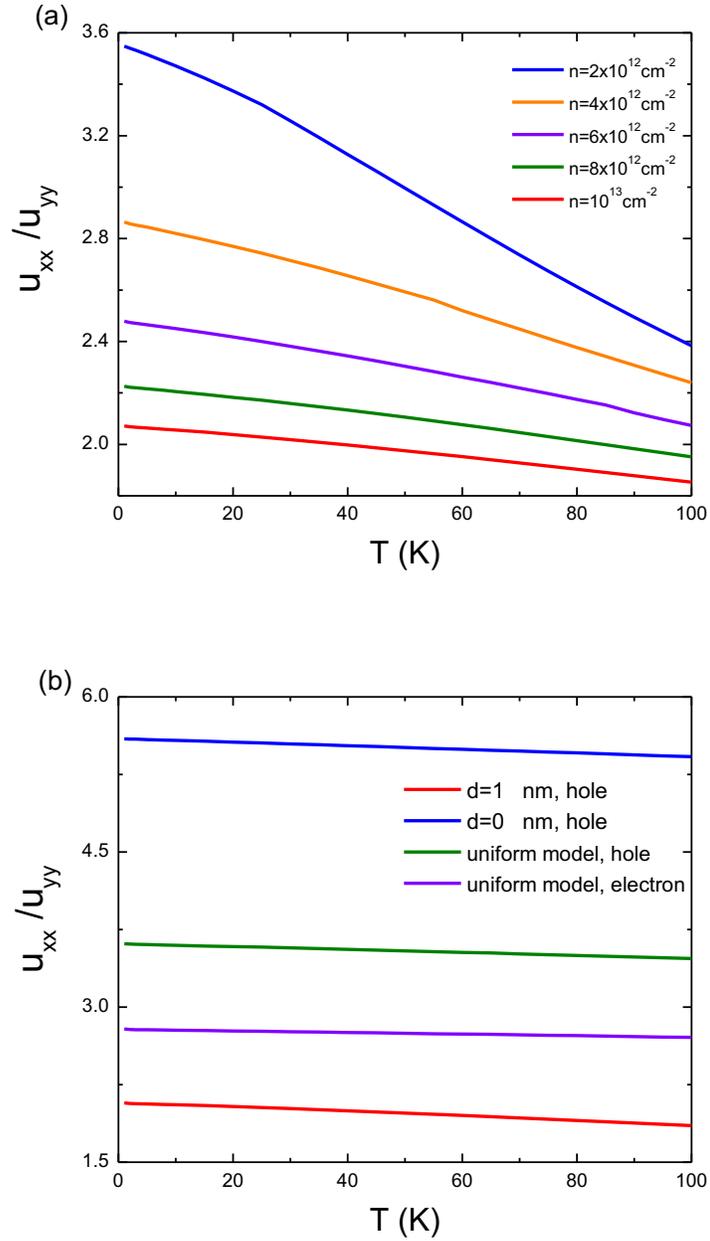

Fig. 5. (Color online) Anisotropy ratio, $\frac{\mu_{xx}}{\mu_{yy}}$, as a function of temperature (a) Hole mobility, $d = 1nm$ and different carrier densities. (b) Fixed carrier density, $n = 10^{13} cm^{-2}$, and $d = 0, d = 1nm$, as well as results for a uniform distribution of impurities for both hole and electron mobility anisotropy ratios.